\documentclass[preprint,aps,preprintnumbers,nofootinbib,superscriptaddress,byrevtex,showpacs]{revtex4}
\usepackage{amsmath,amssymb,amsbsy,mathbbol}
\usepackage{graphicx,subfigure}
\usepackage{comment}

\begin{document}

\unitlength=1mm

\def\a{{\alpha}}
\def\b{{\beta}}
\def\d{{\delta}}
\def\D{{\Delta}}
\def\e{{\epsilon}}
\def\g{{\gamma}}
\def\G{{\Gamma}}
\def\k{{\kappa}}
\def\l{{\lambda}}
\def\L{{\Lambda}}
\def\m{{\mu}}
\def\n{{\nu}}
\def\w{{\omega}}
\def\O{{\Omega}}
\def\S{{\Sigma}}
\def\s{{\sigma}}
\def\t{{\tau}}
\def\th{{\theta}}
\def\x{{\xi}}

\def\ol#1{{\overline{#1}}}

\def\Dslash{D\hskip-0.65em /}
\def\dslash{{\partial\hskip-0.5em /}}
\def\vslash{{\rlap \slash v}}
\def\qbar{{\overline q}}

\def\CPT{{$\chi$PT}}
\def\QCPT{{Q$\chi$PT}}
\def\PQCPT{{PQ$\chi$PT}}
\def\tr{\text{tr}}
\def\str{\text{str}}
\def\diag{\text{diag}}
\def\order{{\mathcal O}}
\def\vit{{\it v}}
\def\vD{\vit\cdot D}
\def\am{\alpha_M}
\def\bm{\beta_M}
\def\gm{\gamma_M}
\def\smb{\sigma_M}
\def\smt{\overline{\sigma}_M}
\def\tb{{\tilde b}}

\def\c#1{{\mathcal #1}}

\def\Bbar{\overline{B}}
\def\Tbar{\overline{T}}
\def\cBbar{\overline{\cal B}}
\def\cTbar{\overline{\cal T}}
\def\pq{(PQ)}

\def\eqref#1{{(\ref{#1})}}

\begin{figure}[!t]
\vskip -1.15cm
\leftline{\includegraphics[width=0.2\textwidth]{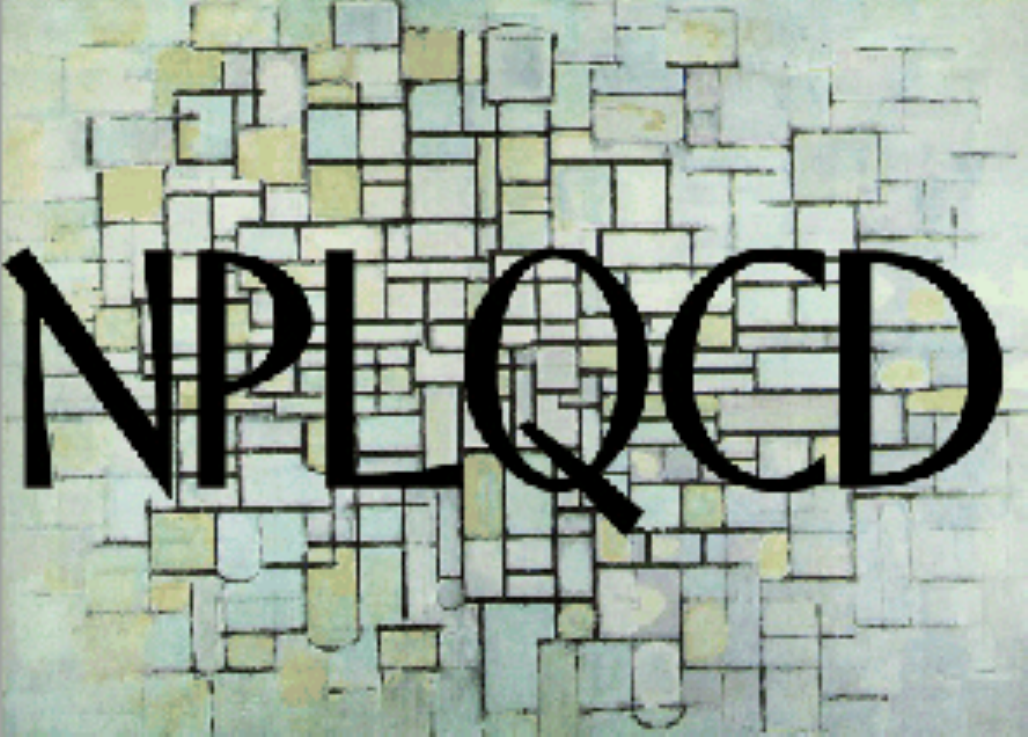}}
\end{figure}

\preprint{NT-UW 05-13}

\preprint{CALT 68-2579}


\title{\bf \large Ginsparg-Wilson Pions Scattering in a Sea of Staggered Quarks}

\author{Jiunn-Wei Chen}
\email[]{jwc@phys.ntu.edu.tw} 
\affiliation{Department of Physics, National Taiwan University, Taipei 10617, Taiwan}

\author{Donal O'Connell}
\email[]{donal@theory.caltech.edu}
\affiliation{California Institute of Technology, Pasadena, CA 91125, USA}

\author{Ruth Van de Water}
\email[]{ruthv@fnal.gov}
\affiliation{Department of Physics, University of Washington,
	Box 351560,
	Seattle, WA 98195-1560, USA}

\author{ Andr\'e Walker-Loud} 
\email[]{walkloud@u.washington.edu}
\affiliation{Department of Physics, University of Washington,
	Box 351560,
	Seattle, WA 98195-1560, USA}

\begin{abstract}
We calculate isospin 2 pion-pion scattering
in chiral perturbation theory for a partially quenched, mixed action
theory with Ginsparg-Wilson valence quarks and staggered sea quarks. 
We point out that for some scattering channels, the power-law volume dependence of two pion states in nonunitary theories such as partially quenched or mixed action QCD is \emph{identical} to that of  QCD.  Thus one can extract  infinite volume scattering parameters from mixed action simulations.  
We then determine the scattering length for both 2 and
2+1 sea quarks in the isospin limit. The scattering length, when expressed
in terms of the pion mass and the decay constant measured on
the lattice, has no contributions from mixed valence-sea mesons, 
thus it does not depend upon the parameter, $C_\textrm{Mix}$, that appears
in the chiral Lagrangian of the mixed theory. 
In addition, the contributions  which nominally arise from operators appearing in the mixed action $\c{O}(a^2 m_q)$ Lagrangian exactly cancel when the scattering length is written in this form.  This is in contrast to the scattering length expressed in terms of the bare parameters of the chiral Lagrangian, which explicitly exhibits all the sicknesses and lattice spacing dependence allowed by a partially quenched mixed action theory.  These results hold for both
2 and 2+1 flavors of sea quarks.
\end{abstract}

\maketitle

%
%
%
%
%
%
%
%
%
%
%
%
\section{Introduction}

Lattice QCD can, in principle, be used to calculate precisely
low energy quantities including hadron masses, decay constants, and
form factors.  In practice, however, limited computing resources make
it currently impossible to calculate processes with dynamical quark masses as
light as those in the real world. Thus one performs simulations with
quark masses that are as light as possible and then extrapolates the
lattice calculations to the physical values using expressions calculated in
chiral perturbation theory ($\chi$PT).  This, of course, relies on the
assumption that the quark masses are light enough that one is in the
chiral regime and can trust $\chi$PT to be a good effective theory of
QCD~\cite{Bernard:2002yk,Beane:2004ks}.

Lattice simulations with staggered fermions~\cite{Susskind:1976jm}
can at present reach significantly lighter quark masses
than other fermion discretizations and have proven extremely
successful in accurately reproducing experimentally measurable
quantities~\cite{Davies:2003ik,Aubin:2004fs}. Staggered fermions,
however, have the disadvantage that each quark flavor comes in
four tastes.  Because these species are degenerate in the continuum,
one can formally remove them by taking the fourth root of the quark
determinant.  In practice, however, the fourth root must be taken
before the continuum limit;  thus it is an open theoretical question
whether or not this fourth-rooted theory becomes QCD in the continuum limit.%
\footnote{See Ref.~\cite{Durr:2005ax}
for a recent review of staggered fermions and the fourth-root trick.}
Even if one assumes the validity of the fourth-root trick, which we do
in the rest of this paper, staggered fermions have other drawbacks.
On the lattice, the four tastes of each quark flavor are no longer
degenerate, and this taste symmetry breaking is numerically
significant in current simulations~\cite{Aubin:2004fs}.  Thus one
must use staggered chiral perturbation theory (S$\chi$PT),
which accounts for taste-breaking discretization effects,
to extrapolate correctly staggered lattice calculations to the
continuum ~\cite{Lee:1999zx,Aubin:2003mg,Aubin:2003uc,Sharpe:2004is}.
Fits of S$\chi$PT expressions for meson masses and decay constants
have been remarkably successful.  Nevertheless, the large number
of operators in the next-to-leading order (NLO) staggered chiral
Lagrangian~\cite{Sharpe:2004is} and the complicated form of the kaon
B-parameter in S$\chi$PT~\cite{VandeWater:2005uq} both show that S$\chi$PT
expressions for many physical quantities will contain a daunting number
of undetermined fit parameters.  Another practical hindrance to the
use of staggered fermions as valence quarks is the construction of
lattice interpolating fields.  Although the construction of a staggered
interpolating field is straightforward for mesons since they are spin 0
objects~\cite{Golterman:1984cy,Golterman:1985dz}, this is not in general
the case for vector mesons, baryons or multi-hadron states since the
lattice rotation operators mix the spin, angular momentum and taste of
a given interpolating field~\cite{Golterman:1984dn,Golterman:1986jf,Bailey:2005ss}.

The use of Ginsparg-Wilson (GW) fermions~\cite{Ginsparg:1981bj} evades both
the practical and theoretical issues associated with staggered fermions.
Because GW fermions are tasteless, one can simply construct
interpolating operators with the right quantum numbers for the desired
meson or baryon.  Moreover, massless GW fermions possess an exact
chiral symmetry on the lattice~\cite{Luscher:1998pq} which protects
expressions in $\chi$PT from becoming unwieldy.\footnote{In practice,
the degree of chiral symmetry is limited by how well the domain-wall
fermion~\cite{Kaplan:1992bt,Shamir:1993zy,Furman:1994ky}
is realized or the overlap
operator~\cite{Narayanan:1993sk,Narayanan:1993ss,Narayanan:1994gw} is
approximated.}  Unfortunately, simulations with dynamical GW quarks are approximately 10
to 100 times slower than those with staggered quarks~\cite{Kennedy:2004ae}
and thus are not presently practical for realizing light quark masses.

A practical compromise is therefore the use of GW valence quarks and
staggered sea quarks.  This so-called ``mixed action" theory
is particularly appealing because the MILC improved staggered
field configurations are publicly available.  Thus one only needs
to calculate correlation functions on top of these background
configurations, making the numerical cost comparable to that
of quenched GW simulations. Several lattice calculations using
domain-wall or overlap valence quarks with the MILC configurations
are underway~\cite{Renner:2004ck,Bowler:2004hs,Bonnet:2004fr},
including a determination of the isospin 2 ($I=2$)
$\pi\pi$ scattering length~\cite{Beane:2005rj}.  Although
this is not the first $I=2$ $\pi\pi$ scattering lattice
simulation~\cite{Sharpe:1992pp,Gupta:1993rn,Aoki:2002in,Yamazaki:2004qb,Aoki:2005uf},
it is the only one with pions light enough to be in the chiral
regime~\cite{Bernard:2002yk,Beane:2004ks}.  Its precision is limited,
however, without the appropriate mixed action $\chi$PT expression for use
in continuum and chiral extrapolation of the lattice data.  With this
motivation we calculate the $I=2$ $\pi\pi$ scattering length in chiral perturbation theory for
a mixed action theory with GW valence quarks and staggered sea quarks.

Mixed action chiral perturbation theory (MA$\chi$PT) was first introduced
in Refs.~\cite{Bar:2002nr,Bar:2003mh,Tiburzi:2005vy} and was extended to include GW
valence quarks on staggered sea quarks for both mesons and baryons
in Refs.~\cite{Bar:2005tu} and~\cite{Tiburzi:2005is}, respectively.
$\pi\pi$ scattering is well understood in continuum, infinite-volume
\CPT~\cite{Weinberg:1966kf,Gasser:1983yg,Gasser:1985gg,Knecht:1995tr,Bijnens:1995yn,Bijnens:1997vq,Bijnens:2004eu},
and is the simplest two-hadron process that one can study
numerically with LQCD.  We extend the NLO \CPT\ calculations
of Refs.~\cite{Gasser:1983yg,Gasser:1985gg} to MA\CPT.
A mixed action simulation necessarily involves partially quenched QCD
(PQQCD)~\cite{Bernard:1993sv,Sharpe:1997by,Golterman:1997st,Sharpe:2000bc,Sharpe:2001fh,Sharpe:2003vy},
in which the valence and sea quarks are treated differently. Consequently,
we provide the PQ\CPT\ $\pi\pi$ scattering amplitude by taking an
appropriate limit of our MA\CPT\ expressions. In all of our computations,
we work in the isospin limit both in the sea and valence sectors.

\bigskip

This article is organized as follows.  We first comment on the determination of infinite volume scattering parameters from lattice simulations in Section~\ref{sec:malqcd}, focusing on the applicability of L\"{u}scher's method~\cite{Luscher:1986pf,Luscher:1990ux} to mixed action lattice simulations.  We then review mixed action LQCD
and MA\CPT\ in Section~\ref{sec:mixed}.  In Section~\ref{sec:mixedScatt}
we calculate the $I=2\ \pi\pi$ scattering amplitude in MA\CPT, first
by reviewing $\pi\pi$ scattering in continuum $SU(2)$ \CPT\ and then
by extending to partially quenched mixed action theories with $N_f=2$
and $N_f=2+1$ sea quarks.  We discuss the role of the double poles in
this process~\cite{Bernard:1993ga} and parameterize the partial quenching
effects in a particularly useful way for taking various interesting and
important limits.  Next, in section~\ref{sec:mixedLength}, we present results for the pion scattering length
in both 2 and $2+1$ flavor MA$\chi$PT.  
These expressions show that it is advantageous to fit to partially
quenched lattice data using the lattice pion mass and pion decay
constant measured on the lattice rather than the LO parameters in the chiral Lagrangian.
We also give expressions for the corresponding continuum PQ$\chi$PT scattering amplitudes, which do not already appear in the literature.    
Finally, in Section~\ref{sec:summary} we briefly discuss how to use our MA$\chi$PT formulae to determine the physical scattering length in QCD from mixed action lattice data and conclude.

%
%
%
%
%
%
%
%
%
%
%
%
\section{Determination of scattering parameters from Mixed Action lattice simulations}\label{sec:malqcd}

Lattice QCD calculations are performed in Euclidean spacetime,
thereby precluding the extraction of S-matrix elements from
infinite volume~\cite{Maiani:1990ca}.  L\"{u}scher, however,
developed a method to extract the scattering phase shifts of two
particle scattering states in quantum field theory by studying the
volume dependence of two-point correlation functions in Euclidean
spacetime~\cite{Luscher:1986pf,Luscher:1990ux}.  In particular, for two particles of equal mass $m$ in an $s$-wave state with zero total 3-momentum in a finite volume, the difference between the energy of the two particles and twice their rest mass is related to the $s$-wave scattering length:\footnote{Here we use
the ``particle physics" definition of the scattering length which is
opposite in sign to the ``nuclear physics" definition.}
\begin{equation}\label{eq:LusForm}
	\Delta E_0 = -\frac{4\pi a_0}{m \textrm{L}^3} \left[ 1 + c_1 \frac{a_0}{\textrm{L}} + c_2 \left(\frac{a_0}{\textrm{L}}\right)^2 + \c{O}\left(\frac{1}{\textrm{L}^3}\right) \right]\,.
\end{equation}
In the above expression, $a_0$ is the scattering length (not to be confused with the lattice spacing, $a$), L is the length of one side of the spatially symmetric lattice, and $c_1$ and $c_2$ are known geometric coefficients.%
\footnote{This expression generalizes to scattering parameters of higher partial waves and non-stationary particles~\cite{Luscher:1986pf,Luscher:1990ux,Rummukainen:1995vs,Kim:2005gf}.}
Thus, even though one cannot directly calculate scattering amplitudes with lattice simulations, 
Eq.~(\ref{eq:LusForm}), which we will refer to as L\"{u}scher's formula, allows one to determine the infinite volume scattering length.  One can then use the expression for the scattering length computed in infinite volume \CPT\ to extrapolate the lattice data to the physical quark masses.  

Because L\"{u}scher's method requires the extraction of energy levels, it relies upon the existence of a Hamiltonian for the theory being studied.  This has not been demonstrated (and is likely false) for partially quenched and mixed action QCD, both of which are nonunitary.  Nevertheless, 
one can calculate the ratio of the two-pion correlator to the square of the single-pion correlator in lattice simulations of these theories and extract the coefficient of the term which is linear in time, which becomes the energy shift in the QCD (and continuum) limit.  We claim that in certain scattering channels, despite the inherent sicknesses of partially quenched and mixed action QCD, this quantity is still related to the infinite volume scattering length via Eq.~(\ref{eq:LusForm}), \emph{i.e.} the volume dependence is identical to Eq.~\eqref{eq:LusForm} up to exponentially suppressed corrections.%
\footnote{Here, and in the following discussion, we restrict ourselves to a perturbative analysis.}
This is what we mean by ``L\"{u}scher's method" for nonunitary theories.  We will expand upon this point in the following paragraphs.

It is well known that L\"{u}scher's formula does not hold for many scattering channels in quenched theories because unitarity-violating diagrams give rise to enhanced finite volume effects~\cite{Bernard:1995ez}.  For certain scattering channels, however,  quenched $\chi$PT calculations in finite volume show that, at 1-loop order, the volume dependence is identical in form to L\"{u}scher's formula~\cite{Bernard:1995ez,Colangelo:1997ch,Lin:2002aj}.  Chiral perturbation theory calculations additionally show that the same sicknesses that generate enhanced finite volume effects in quenched QCD also do so in partially quenched and mixed action theories~%
\cite{Sharpe:2000bc,Sharpe:2001fh,Beane:2002np,Bar:2002nr,Bar:2003mh,Lin:2003tn,Bar:2005tu,Golterman:2005xa}.  It then follows that if a given scattering channel has the same volume dependence as Eq.~\eqref{eq:LusForm} in quenched QCD, the corresponding partially quenched (and mixed action) two-particle process will also obey Eq.~\eqref{eq:LusForm}.  Correspondingly, scattering channels which have enhanced volume dependence in quenched QCD also have enhanced volume dependence in partially quenched and mixed action theories.  We now proceed to discuss in some detail why 
L\"{u}scher's formula does or does not hold for various 2$\rightarrow$2 scattering channels.  

Finite volume effects in lattice simulations come from the ability of particles to propagate over long distances and feel the finite extent of the box through boundary conditions.  Generically, they are proportional either to inverse powers of L or to exp(-$m$L), but L\"{u}scher's formula neglects exponentially suppressed corrections.  Calculations of scattering processes in effective field theories at finite volume show that the power-law corrections only arise from $s$-channel diagrams~%
\cite{Bernard:1995ez,Lin:2002nq,Lin:2002aj,Lin:2003tn,Beane:2003da}.  This is because all of the intermediate particles can go on-shell simultaneously, and thus are most sensitive to boundary effects.  Consequently, when there are no unitarity-violating effects in the $s$-channel diagrams for a particular scattering process, the volume dependence will be identical to Eq.~\eqref{eq:LusForm}, up to exponential corrections.  Unitarity-violating \emph{hairpin} propagators in $s$-channel diagrams, however,  give rise to enhanced volume corrections because they contain double poles which are more sensitive to boundary effects~\cite{Bernard:1995ez}.%
\footnote{We note that, while the enhanced volume corrections in quenched QCD invalidate the extraction of scattering parameters from certain scattering channels, \emph{e.g.} $I=0$~\cite{Bernard:1995ez,Lin:2002aj}, this is not the case in principle for partially quenched QCD, since QCD is a subset of the theory.  Because the enhanced volume contributions must vanish in the QCD limit, they provide a ``handle" on the enhanced volume terms.  In practice, however, these enhanced volume terms may dominate the correlation function, making the extraction of the desired (non-enhanced) volume dependence impractical.}
Thus all violations of L\"{u}scher's formula come from on-shell hairpins in the $s$-channel.  

Let us now consider $I=2$ $\pi\pi$ scattering in the mixed action theory.  
All intermediate states must have isospin 2 and $s\geq 4m^2$.  If one cuts an arbitrary graph connecting the incoming and outgoing pions, there is only enough energy for two of the internal pions to be on-shell, and, by conservation of isospin, they must be valence $\pi^+$'s.\footnote{We restrict the incoming pions to be below the inelastic threshold;  this is necessary for the validity of L\"{u}scher's formula even in QCD.}  Thus no hairpin diagrams ever go on-shell in the $s$-channel, and the structure of the integrals which contribute to the power-law volume dependence in the partially quenched and mixed action theories is identical to that in continuum $\chi$PT.   This insures that L\"uscher's formula is correctly reproduced to all orders in 1/L with the correct ratios between coefficients of the various terms.  Moreover, this holds to all orders in $\chi$PT, PQ$\chi$PT, MA$\chi$PT, and even quenched $\chi$PT.  The sicknesses of the partially quenched and mixed action theories only alter the exponential volume dependence of the $I=2$ scattering amplitude.%
\footnote{In fact, hairpin propagators will give larger exponential dependence than standard propagators because they are more chirally sensitive.}
This is in contrast to the $I=0$ $\pi\pi$ amplitude, which suffers from enhanced volume corrections away from the QCD limit.  In general, the argument which protects L\"{u}scher's formula from enhanced power-like volume corrections holds for all ``maximally-stretched" states at threshold in the meson sector, i.e. those with the maximal values of all conserved quantum numbers;  other examples include $K^+K^+$ and $K^+\pi^+$ scattering.  We expect that a similar argument will hold for certain scattering channels in the baryon sector.  

Therefore the $s$-wave $I=2$ $\pi\pi$ scattering length can be extracted from mixed action lattice simulations using L\"{u}scher's formula and then extrapolated to the physical quark masses and to the continuum using the infinite volume MA$\chi$PT expression for the scattering length.%
\footnote{For a related discussion, see Ref.~\cite{Bedaque:2004ax}}

%
%
%
%
%
%
%
%
%
%
%
%
\section{Mixed Action Lagrangian and Partial Quenching}\label{sec:mixed}

Mixed action theories use different discretization techniques in
the valence and sea sectors and are therefore a natural extension
of partially quenched theories.  We consider a theory with
$N_f$ staggered sea quarks and $N_v$ valence quarks (with $N_v$
corresponding ghost quarks) which satisfy the Ginsparg-Wilson
relation~\cite{Ginsparg:1981bj,Luscher:1998pq}.  In particular we
are interested in theories with two light dynamical quarks ($N_f
= 2$) and with three dynamical quarks where the two light quarks are degenerate (commonly
referred to as $N_f = 2+1$).  To construct the continuum effective
Lagrangian which includes lattice artifacts one follows the two
step procedure outlined in Ref.~\cite{Sharpe:1998xm}.  First one
constructs the Symanzik continuum effective Lagrangian at the quark
level~\cite{Symanzik:1983gh,Symanzik:1983dc} up to a given order in the
lattice spacing, $a$:
\begin{equation}
	\c{L}_{Sym} = \c{L} + a \c{L}^{(5)} + a^2 \c{L}^{(6)} + \ldots,
\end{equation}
where $\c{L}^{(4+n)}$ contains higher dimensional operators of dimension
$4+n$. Next one uses the method of spurion analysis to map the Symanzik
action onto a chiral Lagrangian, in terms of pseudo-Goldstone mesons,
which now incorporates the lattice spacing effects. This has been done
in detail for a mixed GW-staggered theory in Ref.~\cite{Bar:2005tu};
here we only describe the results.

The leading quark level Lagrangian is given by
\begin{equation}\label{eq:LOqLag}
	\c{L} = \sum_{a,b=1}^{4N_f +2N_v} 
		\bar{Q}^{a} \left[ i \Dslash -m_Q \right]_a^{\ b} Q_b,
\end{equation}
where the quark fields are collected in the vectors
\begin{align}
	Q^{N_f = 2} &= (\underbrace{u,d}_\textrm{valence},
		\underbrace{j_1, j_2, j_3, j_4, l_1, l_2, l_3, l_4}_\textrm{sea}, 
		\underbrace{\tilde{u},\tilde{d}}_\textrm{ghost})^T, \\
	Q^{N_f = 2+1} &= (\underbrace{u,d,s}_\textrm{valence}, 
		\underbrace{j_1,j_2,j_3,j_4, l_1,l_2,l_3,l_4, r_1,r_2,r_3,r_4}_\textrm{sea}, 
		\underbrace{\tilde{u},\tilde{d},\tilde{s}}_{\textrm{ghost}})^T
\end{align}
for the two theories. There are 4 tastes for each flavor of sea
quark, $j,l,r$.\footnote{Note that we use different labels for the
valence and sea quarks than Ref.~\cite{Bar:2005tu}.  Instead we use
the ``nuclear physics'' labeling convention, which is consistent with
Ref.~\cite{Tiburzi:2005is}.}  We work in the isospin limit in both the
valence and sea sectors so the quark mass matrix in the 2+1 sea flavor
theory is given by
\begin{equation}\label{eq:Masses}
	m_Q = \text{diag}(\underbrace{m_u, m_u, m_s}_\textrm{valence}, 
			\underbrace{m_j, m_j, m_j, m_j, m_j, m_j, m_j, m_j, m_r, m_r, m_r, m_r}_\textrm{sea}, 
			\underbrace{m_u, m_u, m_s}_\textrm{ghost}).
\end{equation}
The quark mass matrix
in the two flavor theory is analogous but without strange valence,
sea and ghost quark masses. The leading order mixed action Lagrangian,
Eq.~\eqref{eq:LOqLag}, has an approximate graded chiral symmetry,
$SU(4N_f+N_v|N_v)_L~\otimes~SU(4N_f+N_v|N_v)_R$, which is exact in the
massless limit.~\footnote{This is a ``fake" symmetry of PQQCD. However,
it gives the correct Ward identities and thus can be used to understand
the symmetries and symmetry breaking of PQQCD~\cite{Sharpe:2001fh}.} In
analogy to QCD, we assume that the vacuum spontaneously breaks this
symmetry down to its vector subgroup, $SU(4N_f +N_v | N_v)_V$, giving
rise to $(4N_f +2N_v)^2 -1$ pseudo-Goldstone mesons. These mesons are
contained in the field
\begin{equation}
	\S = {\rm exp} \left( \frac{2 i \Phi}{f} \right) ,  \;\;\; 
	\Phi = \begin{pmatrix}
			M & \chi^\dagger\\
			\chi & \tilde{M}\\
		\end{pmatrix}.
\label{eq:sigma}
\end{equation}
The matrices $M$ and $\tilde{M}$ contain bosonic mesons while $\chi$
and $\chi^\dagger$ contain fermionic mesons. Specifically,
\begin{align}\label{eq:mesons}
	M &=\begin{pmatrix}
			\eta_u & \pi^+ & \ldots & \phi_{uj} & \phi_{ul} & \ldots \\
			\pi^- & \eta_d & \ldots & \phi_{dj} & \phi_{dl} & \ldots \\
			\vdots & \vdots & \ddots & \ldots & \ldots & \ldots \\
			\phi_{ju} & \phi_{jd} & \vdots & \eta_j & \phi_{jl} & \ldots \\
			\phi_{lu} & \phi_{ld} & \vdots & \phi_{lj} & \eta_l & \ldots \\
			\vdots & \vdots & \vdots & \vdots & \vdots & \ddots \\
		\end{pmatrix}\quad ,\quad 
	\tilde M = \begin{pmatrix}
			{\tilde \eta}_u & {\tilde \pi}^+ & \ldots \\
			{\tilde \pi}^- & {\tilde \eta}_d & \ldots \\
			\vdots & \vdots & \ddots \\
		\end{pmatrix}
	\nonumber\\
	\chi &= \begin{pmatrix}
		\phi_{\tilde{u} u} & \phi_{\tilde{u} d} & \ldots & \phi_{\tilde{u} j} & \phi_{\tilde{u} l} & \ldots \\
		\phi_{\tilde{d} u} & \phi_{\tilde{d} d} & \ldots & \phi_{\tilde{d} j} & \phi_{\tilde{d} l} & \ldots \\
		\vdots & \vdots & \vdots & \vdots & \vdots & \ddots \\
		\end{pmatrix}.
\end{align}
In Eq.~\eqref{eq:mesons} we only explicitly show the mesons needed in
the two flavor theory. The ellipses indicate mesons containing strange
quarks in the 2+1 theory. The upper $N_v \times N_v$ block of $M$
contains the usual mesons composed of a valence quark and anti-quark.
The fields composed of one valence quark and one sea anti-quark, such as
$\phi_{uj}$, are $1 \times 4$ matrices of fields where we have suppressed
the taste index on the sea quarks. Likewise, the sea-sea mesons such as
$\phi_{jl}$ are $4 \times 4$ matrix-fields. Under chiral transformations,
$\Sigma$ transforms as
\begin{equation}
	\S \longrightarrow L\ \S\ R^\dagger \quad ,\quad
		L,R \in SU(4N_f +N_v |N_v)_{L,R}.
\end{equation}

In order to construct the chiral Lagrangian it is useful to first define
a power-counting scheme. Continuum \CPT\ is an expansion in
powers of the pseudo-Goldstone meson momentum and mass squared
\cite{Gasser:1983yg,Gasser:1985gg}:
\begin{equation}\label{eq:smallScale}
	\varepsilon^2 \sim p_\pi^2 / \Lambda_\chi^2 \sim m_\pi^2 / \Lambda_\chi^2\,,
\end{equation}
where $m_\pi^2 \propto m_Q$ and $\Lambda_\chi$ is the cutoff of $\chi$PT.
In a mixed theory (or any theory which incorporates lattice spacing
artifacts) one must also include the lattice spacing in the power
counting. Both the chiral symmetry of the Ginsparg-Wilson valence
quarks and the remnant $U(1)_A$ symmetry of the staggered sea quarks
forbid operators of dimension five; therefore the leading lattice spacing
correction for this mixed action theory arises at $\c{O}(a^2)$. Moreover,
current staggered lattice simulations indicate that taste-breaking
effects (which are of $\c{O}(a^2)$) are numerically of the same size
as the lightest staggered meson mass~\cite{Aubin:2004fs}. We therefore
adopt the following power-counting scheme:
\begin{equation}
 	\varepsilon^2 \sim p_\pi^2 / \Lambda_\chi^2 \sim m_Q / \Lambda_\textrm{QCD} 
		\sim a^2 \Lambda_\textrm{QCD}^2\,.
\label{eq:epsilons}
\end{equation}
The leading order (LO), $\c{O}(\varepsilon^2)$, Lagrangian is then given
in Minkowski space by~\cite{Bar:2005tu}
\begin{equation}
	\c{L} = \frac{f^2}{8} \text{str} \left( \partial_\mu \S\, \partial^\mu \S^\dagger \right)
		+ \frac{f^2 B}{4} \text{str} \left( \S m_Q^\dagger + m_Q \S^\dagger \right)
		- a^2 \left( \c{U}_S +\c{U}^\prime_S + \c{U}_V \right),
\end{equation}
where we use the normalization $f \sim 132$~MeV and have already
integrated out the taste singlet $\Phi_0$ field, which is proportional
to str$(\Phi)$~\cite{Sharpe:2001fh}. $\c{U}_S$ and $\c{U}^\prime_S$ are
the well-known taste breaking potential arising from the staggered sea
quarks~\cite{Lee:1999zx,Aubin:2003mg}. The staggered
potential only enters into our calculation through an additive shift to the
sea-sea meson masses; we therefore do not write out its explicit form.
The enhanced chiral properties of the mixed action theory are illustrated
by the fact that only one new potential term arises at this order:
\begin{equation}\label{eq:MixPotential}
	\c{U}_V = -C_\textrm{Mix}\ \text{str} \left( \t_3 \S \t_3 \S^\dagger \right),
\end{equation}
where
\begin{equation}
	\t_3 = \c{P}_S -\c{P}_V = \text{diag}(-I_V, I_t \otimes I_S, -I_V).
\end{equation}
The projectors, $\c{P}_S$ and $\c{P}_V$, project onto the sea and
valence-ghost sectors of the theory, $I_V$ and $I_S$ are the valence and
sea flavor identities, and $I_t$ is the taste identity matrix. From this
Lagrangian, one can compute the LO masses of the various pseudo-Goldstone
mesons in Eq.~\eqref{eq:mesons}. For mesons composed of only valence
(ghost) quarks of flavors $a$ and $b$,
\begin{equation}
	m_{ab}^2 = B (m_a +m_b).
\label{eq:m_tree}\end{equation}
This is identical to the continuum LO meson mass because the chiral
properties of Ginsparg-Wilson quarks protect mesons composed of only
valence (ghost) quarks from receiving mass corrections proportional
to the lattice spacing. However, mesons composed of only sea quarks
of flavors $s_1$ and $s_2$ and taste $t$, or mixed mesons with one
valence ($v$) and one sea ($s$) quark both receive lattice spacing mass
shifts.  Their LO masses are given by
\begin{align}\label{eq:ssvsmasses}
	\tilde{m}_{s_1 s_2,t}^2 &= B(m_{s_1} +m_{s_2}) +a^2 \D(\xi_t), \\
	\tilde{m}_{vs}^2 &= B(m_v +m_s) +a^2 \D_\textrm{Mix}.
\end{align}
From now on we use tildes to indicate masses that include
lattice spacing shifts.  The only sea-sea mesons that enter $\pi\pi$ scattering to the order at
which we are working are the taste-singlet mesons (this is because the
valence-valence pions that are being scattered are tasteless), which are
the heaviest; we therefore drop the taste label, $t$.  The splittings
between meson masses of different tastes have been determined numerically
on the MILC configurations~\cite{Aubin:2004fs}, so $\D(\xi_I)$ should
be considered an input rather than a fit parameter. The mixed mesons
all receive the same $a^2$ shift given by
\begin{equation}
	\D_\textrm{Mix} = \frac{16 C_\textrm{Mix}}{f^2}\,,
\end{equation}
which has yet to be determined numerically.

After integrating out the $\Phi_0$ field, the two point correlation
functions for the flavor-neutral states deviate from the simple single
pole form. The momentum space propagator between two flavor neutral
states is found to be at leading order~\cite{Sharpe:2001fh}
\begin{equation}\label{eq:etaProp}
	\c{G}_{\eta_a \eta_b}(p^2) =
		\frac{i \e_a \d_{ab}}{p^2 -m_{\eta_a}^2 +i\e}
		- \frac{i}{N_f} \frac{\prod_{k=1}^{N_f}(p^2 -\tilde{m}_{k}^2 +i\e)}
			{(p^2 -m_{\eta_a}^2 +i\e)(p^2 -m_{\eta_b}^2 +i\e) \prod_{k^\prime=1}^{N_f-1}
				(p^2 -\tilde{m}_{k^\prime}^2 +i\e)},
\end{equation}
where
\begin{equation}
	\e_a = \left\{ \begin{array}{ll}
			+1& \text{for a = valence or sea quarks}\\
			-1 & \text{for a = ghost quarks}\,.
			\end{array} \right.
\end{equation}
In Eq.~\eqref{eq:etaProp}, $k$ runs over the flavor neutral states
($\phi_{jj}, \phi_{ll}, \phi_{rr}$) and $k^\prime$ runs over the
mass eigenstates of the sea sector. For $\pi\pi$ scattering, it will
be useful to work with linear combinations of these $\eta_a$ fields.
In particular we form the linear combinations
\begin{equation}
	\pi^0 = \frac{1}{\sqrt{2}} \left( \eta_u - \eta_d \right)\quad , \quad
	\bar{\eta} = \frac{1}{\sqrt{2}} \left( \eta_u +\eta_d \right),
\end{equation}
for which the propagators are
\begin{eqnarray}
	\c{G}_{\pi^0}(p^2) &=& \frac{i}{p^2 -m_\pi^2 +i\e},\label{eq:PionProp} \\
	\c{G}_{\bar{\eta}}(p^2) &=& \frac{i}{p^2 -m_\pi^2 +i\e}
		- \frac{2i}{N_f} \frac{\prod_{k=1}^{N_f}(p^2 -\tilde{m}_{k}^2 +i\e)}
			{(p^2 -m_\pi^2 +i\e)^2 \prod_{k^\prime=1}^{N_f-1}
				(p^2 -\tilde{m}_{k^\prime}^2 +i\e)}\label{eq:EtaBarProp}.
\end{eqnarray}
Specifically,
\begin{align}
	\c{G}_{\bar{\eta}}(p^2) &= 
		\frac{i}{p^2 -m_\pi^2} -i \frac{p^2 -\tilde{m}_{jj}^2}{(p^2 -m_\pi^2)^2}, 
			 &\text{for $N_f = 2$}, \\
		&= 
		\frac{i}{p^2 - m_\pi^2} -\frac{2i}{3} \frac{(p^2 - \tilde m_{jj}^2)(p^2 - \tilde m_{rr}^2)}
				{(p^2 - m_\pi^2)^2\, (p^2 - \tilde m_{\eta}^2)},
			&\text{for $N_f = 2+1$,} \label{eq:EtaBarProp63}
\end{align}
where $\tilde{m}_{\eta}^2 = \frac{1}{3}(\tilde{m}_{jj}^2 +2\tilde{m}_{rr}^2)$.

%
%
%
%
%
%
%
%
%
%
%
%
\section{Calculation of the $I=2$ Pion Scattering Amplitude}\label{sec:mixedScatt}

Our goal in this work is to calculate the $I=2$ $\pi\pi$ scattering
length in chiral perturbation theory for a partially quenched, mixed
action theory with GW valence quarks and staggered sea quarks, in order
to allow correct continuum and chiral extrapolation of mixed action
lattice data. We begin, however, by reviewing the pion scattering
amplitude in continuum $SU(2)$ chiral perturbation theory. We next
calculate the scattering amplitude in $N_f = 2$ PQ$\chi$PT and
MA$\chi$PT, and finally in $N_f = 2+1$ PQ$\chi$PT and MA$\chi$PT.
When renormalizing divergent 1-loop integrals, we use dimensional regularization 
and a modified minimal subtraction scheme where we consistently subtract all terms proportional 
to~\cite{Gasser:1983yg}:
\begin{equation*}
\frac{2}{4-d} -\gamma_E + \log 4\pi +1,
\end{equation*} 
where $d$ is the number of space-time dimensions.  The scattering amplitude can
be related to the scattering length and other scattering parameters,
as we discuss in Section~\ref{sec:mixedLength}.

%
%
%
%
%
%
%
%
%
%
%
%
\subsection{Continuum $SU(2)$}

\begin{figure}
\begin{tabular}{ccccc}
	\includegraphics[width=0.28\textwidth]{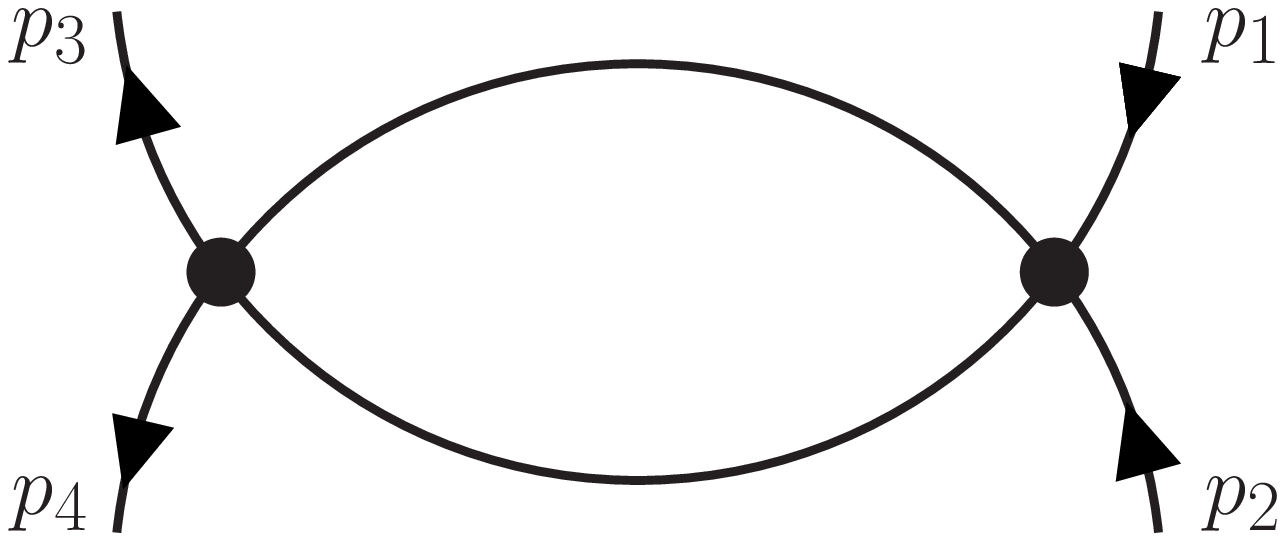} & $\;\;\;\;$ & \includegraphics[width=0.28\textwidth]{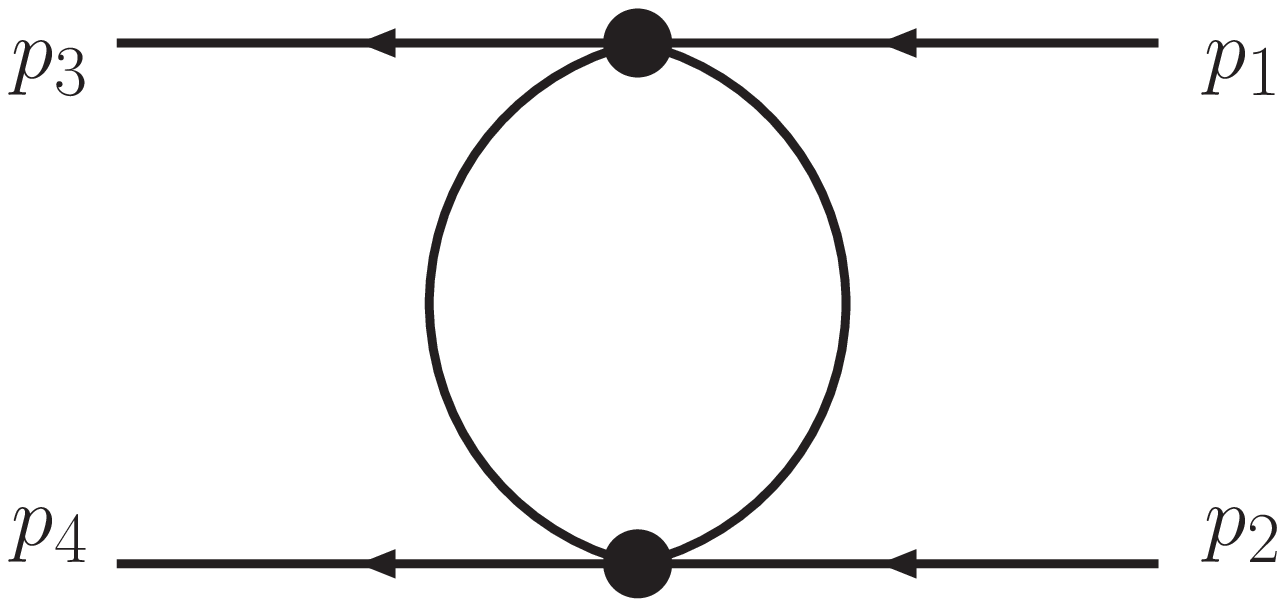} & $\;\;\;\;$ & \includegraphics[width=0.28\textwidth]{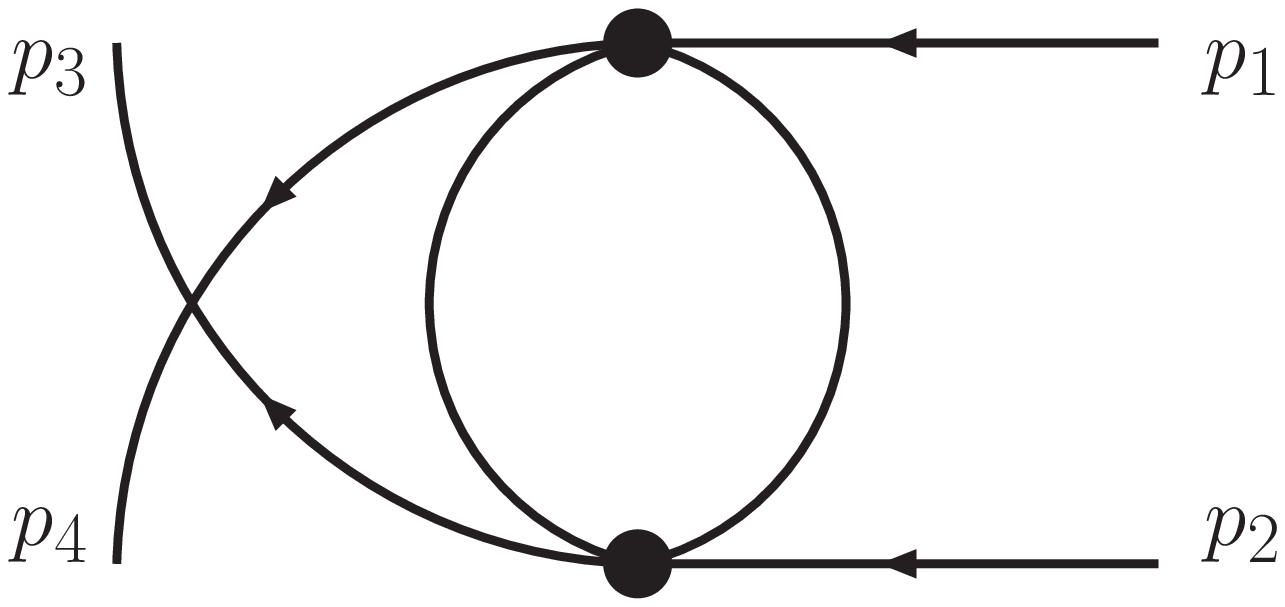} \\
	(a) & $\;\;$ & (b) & $\;\;$ & (c) \\
\end{tabular}
\vspace{2mm}
\begin{tabular}{cccc}
	\includegraphics[width=0.19\textwidth]{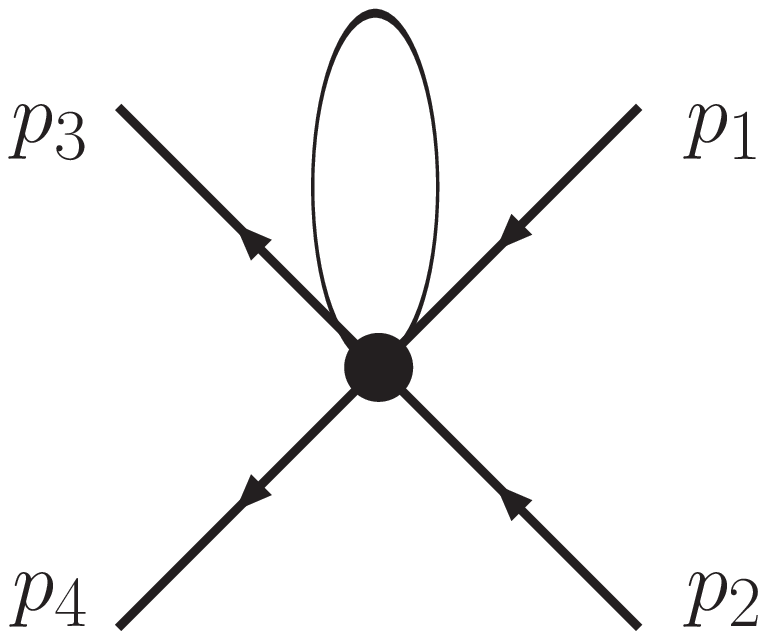} & $\;\;\;\;\;\;$ & \includegraphics[width=0.19\textwidth]{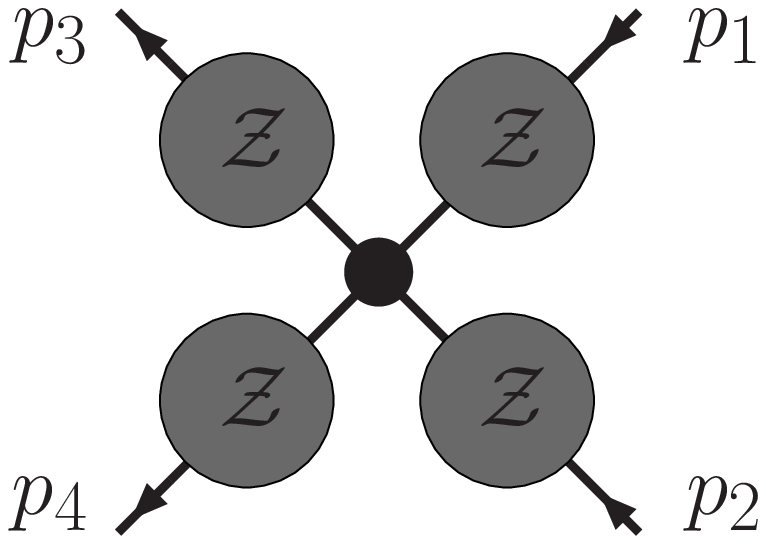}  \\
	(d) & $\;\;\;\;\;\;\;\;\;\;\;\;\;\;$ & (e)  \\
	\end{tabular}
	\caption{One-loop diagrams contributing to the $\pi\pi$ scattering amplitude.  Diagrams (a)--(c) are the $s$-, $t$-, and $u$-channel diagrams, respectively, while diagram (e) represents wavefunction renormalization.}\label{fig:one-loop}
\end{figure}

The tree-level $I=2$ pion scattering amplitude at threshold is
well known to be~\cite{Weinberg:1966kf}
\begin{equation}
i \c{A} = - \frac{4 i m_\pi^2}{f_\pi^2} .
\end{equation}
It is corrected at $\c{O}({\varepsilon^4})$ by loop diagrams and
also by tree level terms from the NLO (or Gasser-Leutwyler)
chiral Lagrangian~\cite{Gasser:1983yg}.%
\footnote{The
continuum $\pi\pi$ scattering amplitude is known to
two-loops~\cite{Knecht:1995tr,Bijnens:1995yn,Bijnens:1997vq,Bijnens:2004eu}.}
The diagrams that
contribute at one loop order are shown in Figure~\ref{fig:one-loop};
they lead to the following NLO expression for the scattering amplitude:
\begin{equation}
        i\c{A}_{\vec{p_i}=0} = -\frac{4i m_{uu}^2}{f^2} \Bigg\{ 1 
                +\frac{m_{uu}^2}{(4\pi f)^2} \bigg[
                        8 \ln \left( \frac{m_{uu}^2}{\mu^2} \right) -1 + l^\prime_{\pi\pi}(\mu) \bigg] \Bigg\} ,
\label{eq:su2amplbare}
\end{equation}
where $m_{uu}$ is the tree-level expression given in
Eq.~(\ref{eq:m_tree}) and $f$ is the LO pion decay constant which appears
in Eq.~(\ref{eq:sigma}).  The coefficient $l^\prime_{\pi\pi}$ is a linear
combination of low energy constants appearing in the Gasser-Leutwyler
Lagrangian whose scale dependence exactly cancels the scale dependence
of the logarithmic term. One can re-express the amplitude, however,
in terms of the physical pion mass and decay constant using the NLO
formulae for $m_\pi$ and $f_\pi$ to find:
\begin{align}
        i\c{A}_{\vec{p_i}=0} &= -\frac{4i m_\pi^2}{f_\pi^2} \Bigg\{ 1 
                +\frac{m_\pi^2}{(4\pi f_\pi)^2} \bigg[ 
                        3 \ln \left( \frac{m_\pi^2}{\mu^2} \right) - 1 +l_{\pi\pi}(\mu) \bigg] \Bigg\},
\label{eq:su2ampl}
\end{align}
where $l_{\pi \pi}$ is a different linear combination of
low energy constants. The expression for $l_{\pi\pi}$ can
be found in Ref.~\cite{Bijnens:1997vq}. We do not, however, include
it here because we do not envision either using the known values of the
Gasser-Leutwyler parameters in the the fit of the scattering length or
using the fit to determine them. The simple expression~\eqref{eq:su2ampl}
has already been used in extrapolation of lattice data from mixed action
simulations~\cite{Beane:2005rj}, but it neglects lattice spacing effects
from the staggered sea quarks which are known from other simulations to be
of the same order as the leading order terms in the chiral expansion of
some observables~\cite{Aubin:2004fs}. We therefore proceed to calculate
the scattering amplitude in a partially quenched, mixed action theory
relevant to simulations.

%
%
%
%
%
%
%
%
%
%
%
%
\subsection{Mixed GW-Staggered Theory with two Sea Quarks}

The scattering amplitude in the partially quenched theory differs
from the unquenched theory in three important respects. First, more
mesons propagate in the loop diagrams. Second, some of the
mesons have more complicated propagators due to hairpin diagrams at
the quark level~\cite{Bernard:1993ga,Sharpe:2001fh}. Third, there are
additional terms in the NLO Lagrangian which arise from partial quenching~\cite{Sharpe:2003vy}, and lattice spacing effects~\cite{Bar:2005tu,Sharpe:2004is}.

\begin{figure}[t]
\centering
\includegraphics[width=0.4\textwidth]{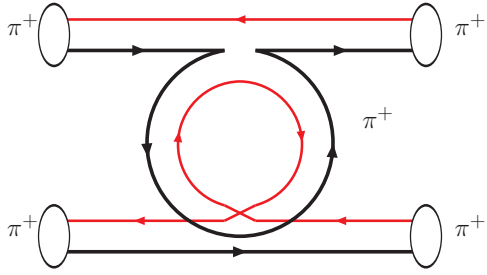}
\caption{Example quark flow for a one-loop $t$-channel graph. This diagram illustrates the presence of meson loops composed of purely valence-valence mesons which are not canceled by valence-ghost loops. Different colors (shades of grey) represent different quark flavors.} 
\label{fig:crossedline}
\end{figure}

At the level of quark flow, there are diagrams such as
Figure~\ref{fig:crossedline}, which route the valence quarks through
the diagram in a way which has no ghostly counterpart. Consequently,
the ghosts do not exactly cancel the valence quarks in loops. Of
course, this is simply a reflection of the fact that the initial and
final states --- valence pions --- are themselves not symmetric under
the interchange of ghost and valence quarks, and therefore the graded
symmetry between the valence and ghost pions has already been violated.
This is well known in quenched and partially quenched heavy baryon
\CPT~\cite{Labrenz:1996jy,Chen:2001yi,Beane:2002vq}.  This fact also
partly explains the success of quenched $\pi\pi$ scattering in the
$I=2$ channel~\cite{Sharpe:1992pp,Gupta:1993rn};  quenching does not
eliminate \emph{all} loop graphs like it does in many other processes,
and in particular, the $s$-channel diagram is not modified by (partial)
quenching effects.  As a consequence, it is necessary to compute all
the graphs contributing to this process in order to determine the scattering
amplitude.

\begin{figure}[t]
\centering
\includegraphics[width=0.28\textwidth]{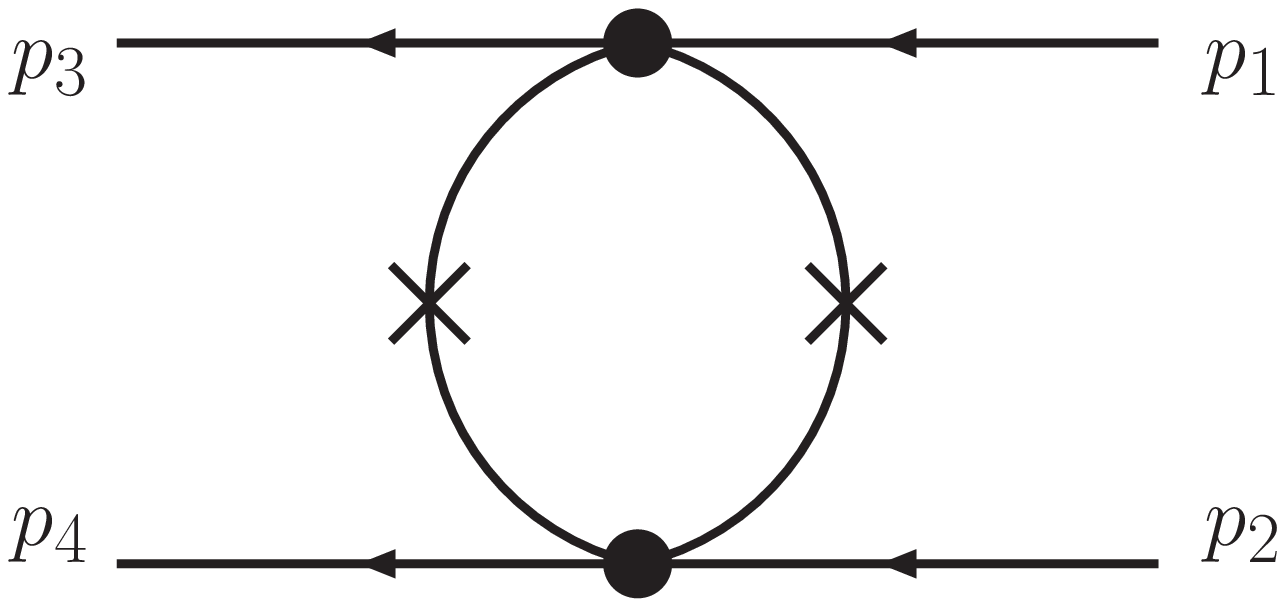}
\hspace{0.05\textwidth}
\includegraphics[width=0.28\textwidth]{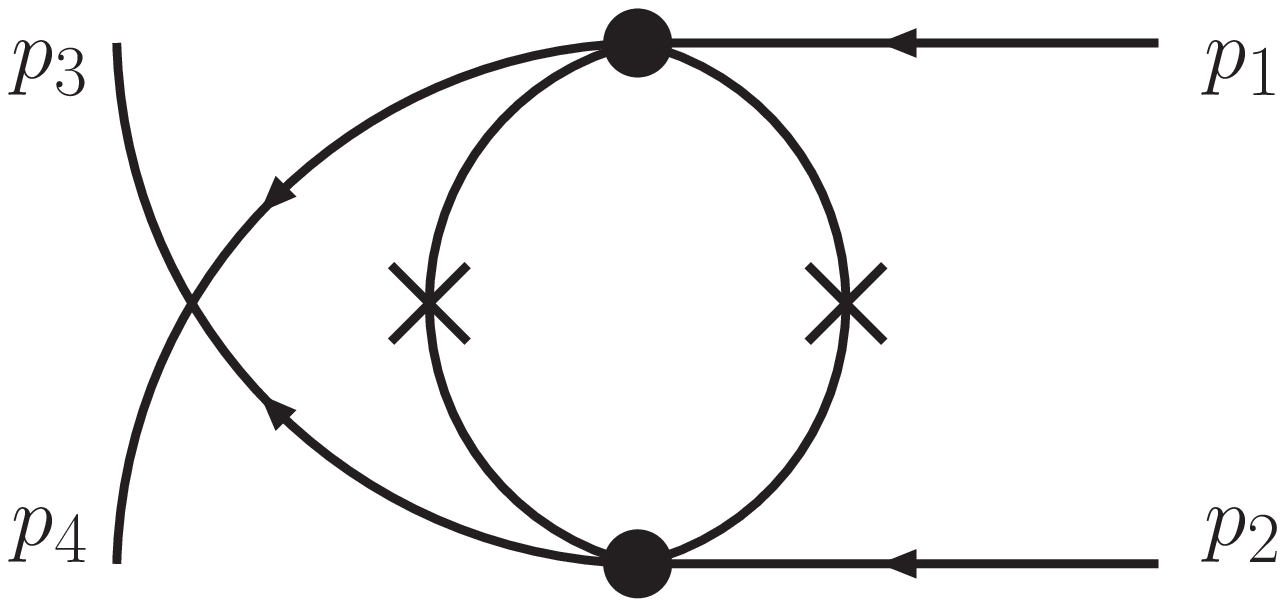}
\caption{Example hairpin diagrams contributing to pion scattering. The propagator
with a cross through it indicates the quark-disconnected piece of the $\bar{\eta}$ propagator, Eq.~\eqref{eq:su42hairprop}.}
\label{fig:Hairpins}
\end{figure}

Quark level disconnected (hairpin) diagrams lead to higher order poles
in the propagator of any particle which has the quantum numbers of the
vacuum~\cite{Bernard:1993ga,Sharpe:2001fh}. In the isospin limit of the
$N_f = 2$ partially quenched theory, conservation of isospin prevents the
$\pi^0$ from suffering any hairpin effects. Hence only the $\bar \eta$
acquires a disconnected propagator. Moreover, in the $m_0 \rightarrow
\infty$ limit, the $\bar \eta$ propagator (given for a general PQ theory
in Eq.~\eqref{eq:EtaBarProp}) is given by the simple expression
\begin{align}
	G_{\bar\eta}(p^2) &= 
		\frac{i}{p^2 -m_\pi^2} -i \frac{p^2 - \tilde m^2_{jj}}{(p^2 - m^2_\pi)^2} \nonumber\\
		&= \frac{i \tilde\D_{PQ}^2}{(p^2 - m_\pi^2)^2},
\label{eq:su42hairprop}
\end{align}
where the parameter
\begin{equation}
        \tilde\D_{PQ}^2 = \tilde{m}_{jj}^2 -m_\pi^2
\end{equation}
quantifies the partial quenching. (Recall that $\tilde{m}_{jj}$ is the
physical mass of a taste \emph{singlet} sea-sea meson.)  Notice that when
$\tilde\D_{PQ} \rightarrow 0$ the propagator~\eqref{eq:su42hairprop} also goes
to zero; this is what we expect since, in the $SU(2)$ theory, the
only neutral propagating state is the $\pi^0$.
The propagator in Eq.~\eqref{eq:su42hairprop} can appear
in loops, thereby producing new diagrams such as those in
Fig.~\ref{fig:Hairpins}.%
\footnote{We note that there are also similar
contributions to the four particle vertex with a loop and to the mass
correction. We do not show them, however, because they cancel against
one another in the amplitude expressed in lattice-physical parameters, which we will show in the following pages.}
After adding all such hairpin diagrams, one finds that the contribution of the $\bar\eta$ to the amplitude is
\footnote{We note that this contribution does not vanish in the limit that $m_\pi^2
\rightarrow 0$ with $\tilde{m}_{jj}^2 \neq 0$.  Similar effects have
been observed in quenched computations of pion scattering
amplitudes~\cite{Colangelo:1997ch,Bernard:1995ez}.  This non-vanishing contribution is the
$I=2$ remnant of the divergences that are known to occur in the $I=0$ amplitude at threshold.  These divergences give rise to enhanced volume corrections to the $I=0$ amplitude with respect to the one-loop $I=2$ amplitude and prevent the use of L\"{u}scher's formula.  Moreover, it is
known~\cite{Sharpe:1997by,Sharpe:2000bc} that PQ\CPT\ is singular in the limit $m_u
\rightarrow 0$ with nonzero sea quark masses, so the behavior of
the amplitude in this limit is meaningless.}
\begin{equation}
	i \c A_{\bar \eta} = 
		\frac{4i}{(4 \pi f_\pi)^2} \frac{\tilde\D_{PQ}^4}{6 f_\pi^2} .
\label{eq:su42amplhairpin}
\end{equation}

In addition to 1-loop contributions, the NLO scattering amplitude receives tree-level analytic contributions from operators of $\c{O}(\epsilon^4)$ in the chiral Lagrangian.  At this order, the mixed action Lagrangian contains the same $\c{O}(p^4)$, $\c{O}(p^2 m_q)$, and $\c{O}(m_q^2)$ operators as in the continuum partially quenched chiral Lagrangian, plus additional $\c{O}(a^4)$, $\c{O}(a^2 m_q)$, and $\c{O}(a^2 p^2)$ operators arising from discretization effects.  We can now enumerate the generic forms of analytic contributions from these NLO operators.  Because of the chiral symmetry of the GW valence sector, all tree-level contributions to the scattering length must vanish in the limit of vanishing valence quark mass.%
\footnote{As we discussed in the previous footnote, this condition need not hold for loop contributions to the scattering amplitude.}
Thus there are only three possible forms, each of which must be multiplied by an undetermined coefficient:  $m^4_{uu}$, $m^2_{uu}m^2_{jj}$, and $m^2_{uu}a^2$.  It may, at first, seem surprising that operators of $\c{O}(a^2m_q)$, which come from taste-symmetry breaking and contain projectors onto the sea sector, can contribute at tree-level to a purely valence quantity.  Nevertheless, this turns out to be the case.  These $\c{O}(a^2m_q)$ mixed action operators can be determined by first starting with the NLO staggered chiral Lagrangian~\cite{Sharpe:2004is}, and then inserting a sea projector, $\c{P}_S$, next to every taste matrix.  One example of such an operator is $\left[\text{\str} \left(\Sigma m_Q^\dagger \right) \text{str}\left( \c{P}_S \xi_5 \Sigma \xi_5 \Sigma^\dagger\right) + \textrm{p.c.}\right]$, where, $\xi_5$ is the $\g_5$ matrix acting in taste-space and p.c. indicates parity-conjugate.  This double-trace operator will contribute to the lattice pion mass, decay constant, and 4-point function at tree-level because one can place all of the valence pions inside the first supertrace, and the second supertrace containing the projector $\c{P}_S$ will just reduce to the identity.  

Putting everything together, the total mixed action scattering amplitude to NLO is
\begin{multline}
        i\c{A}_{\vec{p_i}=0} = -\frac{4i m_{uu}^2}{f^2} \Bigg\{ 1 
                +\frac{m_{uu}^2}{(4\pi f)^2} \Bigg[
                        4 \ln \left( \frac{m_{uu}^2}{\mu^2} \right) 
                +4 \frac{\tilde{m}_{ju}^2}{m_{uu}^2} \ln \left( \frac{\tilde{m}_{ju}^2}{\mu^2} \right) 
                -1 +l^\prime_{\pi\pi}(\mu) \Bigg]
                \\
		- \frac{m_{uu}^2}{(4\pi f)^2} \Bigg[
             	   	\frac{\tilde\D_{PQ}^4}{6 m_{uu}^4}
			+\frac{\tilde\D_{PQ}^2}{m_{uu}^2} \left[ \ln \left( \frac{m_{uu}^2}{\mu^2} \right) +1 \right]
		\Bigg] 
                + \frac{\tilde\D_{PQ}^2}{(4\pi f)^2}l^\prime_{PQ}(\mu) 
                + \frac{a^2}{(4\pi f)^2} l^\prime_{a^2}(\mu)
                \Bigg\}.
\label{eq:2seaBare}
\end{multline}
The first line of Eq.~\eqref{eq:2seaBare} contains those terms which remain in the continuum and full QCD limit, Eq.~\eqref{eq:su2amplbare}, while the second line accounts for the effects of partial quenching and of the nonzero lattice spacing.  Note that, for consistency with the 1-loop terms, we chose to re-express the analytic contribution proportional to the sea quark mass as $m^2_{uu}\tilde\Delta_{PQ}^2$.  In Eq.~\eqref{eq:2seaBare} we have multiplied every contribution
from diagrams which contain a sea quark loop by $1/4$, thus making our expression applicable to lattice simulations in which the fourth root of the staggered sea quark determinant is taken.

It is useful, however, to  re-express the scattering amplitude in terms of the quantities that one measures in a lattice simulation:   $m_\pi$ and $f_\pi$.  Throughout this paper, we will refer to these renormalized measured quantities as the lattice-physical pion mass and decay constant.\footnote{Notice that once the lattice spacing $a$ has been determined, the lattice-physical pion mass can be
unambiguously determined by measuring the exponential decay of a pion-pion correlator. We assume that the lattice spacing $a$ has been determined, for example, by studying the heavy quark potential
or quarkonium spectrum.}   Because we are working consistently to second order in chiral perturbation theory, we can equate the lattice-physical pion mass to the 1-loop chiral perturbation theory expression for the pion mass, and likewise for the lattice-physical decay constant.  Thus, in terms of lattice-physical parameters, the mixed action $I=2$ $\pi\pi$ scattering amplitude is
\begin{equation}
        i\c{A}^{MA{\chi}PT}_{\vec{p_i}=0} = -\frac{4i m_\pi^2}{f_\pi^2} \Bigg\{ 1 
		+ \frac{m_\pi^2}{(4\pi f_\pi)^2} \bigg[ 
			3\ln \left( \frac{m_\pi^2}{\mu^2} \right) 
			-1 +l_{\pi\pi}(\mu) \bigg]
		-\frac{m_\pi^2}{(4\pi f_\pi)^2} \frac{\tilde\D_{PQ}^4}{6\, m_\pi^4} 
	\Bigg\},
\label{eq:su42ampl}
\end{equation}
where the first few terms are identical in form to the full QCD amplitude, Eq.~\eqref{eq:su2ampl}.  This expression for the scattering amplitude is vastly simpler than the one in terms of the bare parameters.  First, the hairpin contributions from all diagrams except those in Fig.~\ref{fig:Hairpins} have exactly cancelled, removing the enhanced chiral logs and leaving the last term in Eq.~\eqref{eq:su42ampl} as the only explicit modification arising from the partial quenching and discretization effects.  Second, 
all contributions from mixed valence-sea mesons in loops have cancelled, thereby removing the new mixed action parameter, $C_\textrm{Mix}$, completely.%
\footnote{Another consequence of the exact cancelation of the loops with mixed valence-sea quarks is that one does not have to implement the ``fourth-root trick'' through this order.}
Third, all tree-level contributions proportional to the sea quark mass have also cancelled from this expression.  And finally, most striking is the fact that an explicit computation of the $\c{O}(a^2 m_q)$ contributions to the amplitude arising from the NLO mixed action Lagrangian show that these effects exactly cancel when the amplitude is expressed in lattice-physical parameters.  This result will be discussed in detail in Ref.~\cite{Chen:noasqd}.  Thus to reiterate, the only  partial quenching and lattice spacing dependence in the amplitude comes from the hairpin diagrams of Fig.~\ref{fig:Hairpins}, which produce contributions proportional to $\tilde{\D}_{PQ}^4 = (m_{jj}^2 + a^2 \D(\xi_I) - m_\pi^2)^2$, where $m^2_{jj}+a^2\D(\xi_1)$ is the mass-squared of the taste-singlet sea-sea meson.  Moreover, we presume that anyone performing a mixed action lattice simulation will separately measure the taste-singlet sea-sea meson mass and use it as an input to fits of other quantities such as the $\pi\pi$ scattering length.  Thus we do not consider it to be an undetermined parameter.

It is now clear that one should fit $\pi\pi$ scattering lattice data in terms of the
lattice-physical pion mass and decay constant
rather than in terms of the LO pion mass and LO decay constant.  By doing this, one eliminates three undetermined fit parameters:  $C_\textrm{Mix}$, $l'_{PQ}$, and $l'_{a^2}$, as well as the enhanced chiral logs.

%
%
%
%
%
%
%
%
%
%
%
%
\subsection{Mixed GW-Staggered Theory with $2+1$ Sea Quarks}

The $2+1$ flavor theory has three additional quarks -- the strange
valence and ghost and strange sea quarks -- which can lead to new
contributions to the scattering amplitude.  Because we only consider
the scattering of valence pions, however, strange valence quarks cannot
appear in this process.  Thus all new contributions to the scattering
amplitude necessarily come only from the sea strange quark, $r$. Because
the $r$ quark is heavier than the other sea quarks there is $SU(3)$
symmetry breaking in the sea. This symmetry breaking only affects the
pion scattering amplitude, expressed in lattice-physical quantities,
through the graphs with internal $\bar{\eta}$ propagators because the
masses of the mixed valence-sea mesons cancel in the final amplitude as
they did in the earlier two flavor case. In addition, the only signature
of partial quenching in the amplitude comes from these same diagrams. It
is therefore worthwhile to investigate the physics of the neutral meson
propagators further.

There are more hairpin graphs in the $2+1$ flavor theory since the
$\eta_s$ may propagate as well as the $\eta_u$ and the $\eta_d$. Because
these mesons mix with one another, the flavor basis is not the most
convenient basis for the computation. Rather, a useful basis of states
is $\pi^0$, $\bar\eta = (\eta_u + \eta_d)/\sqrt 2$ and $\eta_s$. Since
we work in the isospin limit, the $\pi^0$ cannot mix with $\bar \eta$
or $\eta_s$; in addition, there is no vertex between the $\eta_s$
and $\pi^+ \pi^-$ at this order, so we never encounter a propagating
$\eta_s$. Thus all the PQ effects are absorbed into the $\bar \eta$
propagator, which is given by
\begin{equation}
	G_{\bar\eta}(p^2) = \frac{i}{p^2 - m_\pi^2} 
		-\frac{2i}{3} \frac{(p^2 - \tilde m_{jj}^2)(p^2 - \tilde m_{rr}^2)}
			{(p^2 - m_\pi^2)^2\, (p^2 - \tilde m_{\eta}^2)} .
\label{eq:su63hairpinprop}
\end{equation}
In $SU(3)$ chiral perturbation theory, the neutral mesons are the $\pi^0$
and the $\eta_8$. Therefore, in the PQ theory, we know that there will be
a contribution from the $\bar\eta$ graphs that does not result from
partial quenching or $SU(3)$ symmetry breaking.  Therefore the extra PQ
graphs arising from the internal $\bar{\eta}$ fields must not vanish in
the $\tilde{\D}_{PQ} \rightarrow 0$ limit, in contrast to the two flavor
case of Eq.~\eqref{eq:su42amplhairpin}. 

To make this clear, we can re-express the propagator of
Eq.~\eqref{eq:su63hairpinprop} in terms of $\tilde{\D}_{PQ}$ as
\begin{equation}
	G_{\bar\eta}(p^2) = 
		i \left[ \frac{\tilde{\D}_{PQ}^2}{(p^2 - m_\pi^2)^2} 
		+\frac{1}{3} \frac{1}{p^2 - \tilde m_{\eta}^2} 
			\left(1 - \frac{\tilde{\D}_{PQ}^2}{p^2 -m_\pi^2} \right)^2 \right] .
\label{eq:su63hairprop}
\end{equation}
This propagator has a single pole which is independent of
$\tilde{\D}_{PQ}$, as well as higher order poles that are at least
quadratic in $\tilde{\D}_{PQ}$. It is interesting to consider the large
$m_r$ limit of this propagator. In this limit, $\tilde m_\eta^2 \approx
\frac{4}{3} B m_r$ is also large. For momenta that are small compared
to $\tilde m_{\eta}$, the second term of this equation goes to zero in
the large $m_r$ limit, and the $2+1$ flavor propagator reduces to the
2 flavor propagator, Eq.~\eqref{eq:su42hairprop}, as expected.

While the above expression clarifies the $\tilde{\D}_{PQ}$
dependence of the propagator and the large $m_r$ limit, it obscures the
$SU(3)_{sea}$ limit. An equivalent form of the propagator is
\begin{equation}
	G_{\bar\eta}(p^2) = i \left[ \frac{\tilde{\D}_{PQ}^2}{(p^2 - m_\pi^2)^2} 
		+\frac{1}{3} \left(1 + \frac{\tilde{\D}_3^2}
			{p^2-\tilde m_{\eta}^2}\right) \frac{1}{p^2 - m_\pi^2} 
			\left(1 - \frac{\tilde{\D}_{PQ}^2}{p^2 -m_\pi^2} \right) \right] ,
\label{eq:su63hairpropdelta3}
\end{equation}
where the quantity $\tilde{\D}_3 = \sqrt{\tilde{m}_{\eta}^2 -
\tilde{m}_{jj}^2}$ parametrizes the $SU(3)_{sea}$ breaking. When
$\tilde{\D}_3= 0$ this propagator is similar in form to the corresponding
2 flavor propagator, Eq.~\eqref{eq:su42hairprop}, but it has an additional
single pole due to the extra neutral meson in the $SU(3)$ theory.

Having considered the new physics of the hairpin propagator, we can now
calculate the scattering amplitude. For our purposes here, it is most
convenient to express the total $I=2$ $\pi\pi$ scattering amplitude in
terms of $\tilde{\D}_{PQ}$. Just as in the 2-flavor computation, the NLO analytic contributions due to partial quenching and finite lattice spacing effects exactly cancel when the amplitude is expressed in lattice-physical parameters.  All sea quark mass and lattice spacing dependence comes from the hairpin diagrams, which produce terms proportional to powers of $\tilde\Delta_{PQ}$ with known coefficients.
The amplitude is
\begin{multline}
        i\c{A}^{MA{\chi}PT}_{\vec{p_i}=0} = -\frac{4i m_\pi^2}{f_\pi^2} \Bigg\{ 1 
                +\frac{m_\pi^2}{(4\pi f_\pi)^2} \Bigg[
                        3 \ln \left( \frac{m_\pi^2}{\mu^2} \right) -1 
                        +\frac{1}{9}\left[ \ln \left( \frac{\tilde{m}_{{\eta}}^2}{\mu^2} \right) +1 \right]
                        + \bar{l}_{\pi\pi}(\mu) \Bigg]
                         \\
		+\frac{1}{(4\pi f_\pi)^2} \Bigg[
			-\frac{\tilde{\D}_{PQ}^4}{6 m_\pi^2} 
		+m_\pi^2 \sum_{n=1}^4 \left( \frac{\tilde{\D}_{PQ}^2}{m_\pi^2} \right)^n\, 
                        	\c{F}_n \left( \tilde{m}_{\eta}^2/m_\pi^2 \right)
	\Bigg] \Bigg\},
\label{eq:su63ampl}
\end{multline}
where $\tilde{\D}_{PQ}^2 = m_{jj}^2 + a^2 \D(\xi_I) - m_\pi^2$ and
\begin{subequations}
\begin{align}
        \c{F}_1(x) &= -\frac{2}{9(x-1)^2} \left[ 5(x-1) -(3x +2)\ln (x) \right], \\ 
        \c{F}_2(x) &= \frac{2}{3(x-1)^3} \left[ (x-1)(x+3) -(3x +1)\ln(x) \right], \\ 
        \c{F}_3(x) &= \frac{1}{9(x-1)^4} \left[ (x-1) (x^2 -7x -12) +2(7x+2) \ln(x) \right], \\ 
        \c{F}_4(x) &= -\frac{1}{54 (x-1)^5} \left[ (x-1) (x^2 -8x -17) +6(3x+1)\ln(x) \right] .
\end{align}\label{eq:coolFs}\end{subequations}
The functions $\c F_i$ have the property that $\c F_i(x) \rightarrow 0$
in the limit that $x \rightarrow \infty$. Therefore, when
the strange sea quark mass is very large, i.e. $\tilde m_{\eta}^2
/ m_\pi^2 \gg 1$, the $2+1$ flavor amplitude reduces to the 2 flavor amplitude, Eq.~\eqref{eq:su42ampl}, with the exception of terms that can be absorbed into the analytic terms.  The low energy constants have a scale dependence which exactly cancels the scale dependence in the logarithms.  The coefficient $\bar{l}_{\pi\pi}$ is the same linear combination of Gasser-Leutwyler coefficients that appear in the $SU(3)$ scattering amplitude expressed in terms of the physical pion mass and decay constant~\cite{Knecht:1995tr,Bijnens:2004eu}. 

Because the functions $\c F_i$ depend logarithmically on
$x$, the $2+1$ flavor scattering amplitude features enhanced chiral logarithms~\cite{Sharpe:1997by}
that are absent from the 2 flavor amplitude. 
This is a useful observation, as we will now explain. Because there
is a strange quark in nature and its mass is less than the
QCD scale, $\Lambda_{\textrm QCD}$, lattice simulations must
use $2+1$ quark flavors. It is often practical to fix the strange quark mass
at a constant value near its physical value in these simulations. This
circumstance is helpful because, just as $SU(2)$ chiral perturbation
theory is useful to describe nature at scales smaller than the strange
quark mass, the 2 flavor amplitude given in Eq.~\eqref{eq:su42ampl}
can be used to extrapolate $2+1$ flavor lattice data at energy scales smaller than the
strange sea quark mass used in the simulation (provided, of course, there are no strange valence quarks)~\cite{Chen:2002bz}. This
is valid because, at energy scales smaller than the strange quark mass (or actually twice the strange quark mass, since the purely pionic systems have no valence strange quarks),
one can integrate out the strange quark. This is not an approximation,
because all of the effects of the strange quark are absorbed into a
renormalization of the parameters of the chiral Lagrangian.
Moreover, since the 2 flavor amplitude does not exhibit
enhanced chiral logarithms, signatures of partial quenching can be reduced
by extrapolating lattice data with the 2 flavor, rather than the $2+1$
flavor, expression. We note that in this case the effects of the strange quark are absorbed in the coefficients of the analytic terms appearing in Eq.~\eqref{eq:su42ampl}, and thus they are not constant,
but rather depend logarithmically upon the strange sea quark mass.

%
%
%
%
%
%
%
%
%
%
%
%
\section{$I=2$ Pion Scattering Length Results}\label{sec:mixedLength}

In this section we  present our results for the $s$-wave $I=2\ \pi\pi$ scattering
length in the two theories most relevant to current mixed action lattice
simulations: those with GW valence quarks and either $N_f=2$ or $N_f =
2+1$ staggered sea quarks.  We only present results for the scattering length expressed in lattice-physical parameters.  The $s$-wave scattering length is trivially related
to the full scattering amplitude at threshold by an overall prefactor:

\begin{equation}
	a_{l=0}^{(I=2)} = \frac{1}{32 \pi m_\pi}  \c{A}^{I=2}  \bigg|_{\vec{p_i}=0}.
\end{equation}

%
%
%
%
%
%
%
%
%
%
%
%
\subsection{Scattering Length with 2 Sea Quarks}

The $I=2\ \pi\pi$ $s$-wave scattering length in a MA\CPT\ theory with 2 sea quarks is given by
\begin{equation}\label{eq:2seaScattLength}
	{a_{0}^{(2)}}^{MA\chi PT} = -\frac{m_\pi}{8 \pi f_\pi^2} \Bigg\{ 1 
                + \frac{m_\pi^2}{(4\pi f_\pi)^2} \bigg[ 
                        3\ln \left( \frac{m_\pi^2}{\mu^2} \right) 
                        -1 +l_{\pi\pi}(\mu) \bigg]
		-\frac{m_\pi^2}{(4\pi f_\pi)^2} \frac{\tilde\D_{PQ}^4}{6\, m_\pi^4} 
	\Bigg\},
\end{equation}
where $\tilde{\D}_{PQ}^2 = m_{jj}^2 + a^2 \D(\xi_I) - m_\pi^2$.
The first two terms are the result one obtains in $SU(2)$ \CPT~\cite{Bijnens:1997vq} and the last term is the only new effect arising from the partial quenching and mixed action.  All other possible partial quenching terms, enhanced chiral logs and additional linear combinations of the $\c{O}(p^4)$ Gasser-Leutwyler coefficients, exactly cancel when the scattering length is expressed in terms of lattice-physical parameters.  And, most strikingly, the pion mass, decay constant and the 4-point function all receive 
$\c{O}(a^2 m_q)$ corrections from the lattice, but they exactly cancel in the scattering length expressed in terms of the lattice -physical parameters~\cite{Chen:noasqd}.  It is remarkable that the only artifact of the nonzero lattice spacing, $m^2_{jj} + a^2 \D_I$, can be separately determined simply by measuring the exponential fall-off of the taste-singlet sea-sea meson 2-point function.  Thus there are no undetermined fit parameters in the mixed action scattering length expression from either partial quenching or lattice discretization effects;  there is only the unknown continuum coefficient, $l_{\pi\pi}$.  

One can trivially deduce the continuum PQ scattering
length from Eq.~\eqref{eq:2seaScattLength}: simply let $a \rightarrow 0$, reducing $\tilde
m_{jj} \rightarrow m_{jj} = 2 B m_j$ in $\tilde{\D}_{PQ}$, resulting in
\begin{equation}
      {a_{0}^{(2)}}^{PQ{\chi}PT} = -\frac{m_\pi}{8 \pi f_\pi^2} \Bigg\{ 1 
                + \frac{m_\pi^2}{(4\pi f_\pi)^2} \bigg[ 
                        3\ln \left( \frac{m_\pi^2}{\mu^2} \right) 
                        -1 +l_{\pi\pi}(\mu) \bigg]
                        -\frac{\D_{PQ}^4}{6(4\pi m_\pi f_\pi)^2} 
		 \Bigg\} .
\label{eq:PQsu42Length}
\end{equation}

%
%
%
%
%
%
%
%
%
%
%
%
\subsection{Scattering Length with 2+1 Sea Quarks}

The $I=2\ \pi\pi$ $s$-wave scattering length in a MA\CPT\ theory with 2+1 sea quarks is given by
\begin{multline}\label{eq:2P1seaScattLength}
	{a_{0}^{(2)}}^{MA\chi PT} =  -\frac{m_\pi}{8 \pi f_\pi^2} \Bigg\{ 1 
                +\frac{m_\pi^2}{(4\pi f_\pi)^2} \Bigg[
                        3 \ln \left( \frac{m_\pi^2}{\mu^2} \right) -1 
                        +\frac{1}{9}\left[ \ln \left( \frac{\tilde{m}_{{\eta}}^2}{\mu^2} \right) +1 \right]
                        + \bar{l}_{\pi\pi}(\mu) \Bigg]
                         \\
		+\frac{1}{(4\pi f_\pi)^2} \Bigg[
			-\frac{\tilde{\D}_{PQ}^4}{6 m_\pi^2} 
%
		+m_\pi^2 \sum_{n=1}^4 \left( \frac{\tilde{\D}_{PQ}^2}{m_\pi^2} \right)^n\, 
                        	\c{F}_n \left( \tilde{m}_{\eta}^2/m_\pi^2 \right)
	\Bigg] \Bigg\},
\end{multline}
where the functions $\c F_i$ are defined in Eq.~\eqref{eq:coolFs}.  As in the 2-flavor
MA\CPT\  expression, Eq.~\eqref{eq:2seaScattLength}, the only undetermined parameter is the linear combination of Gasser-Leutwyler coefficients, $\bar{l}_{\pi\pi}$, which also appears in the continuum \CPT\ expression.   

We note as an aside that this suppression of lattice spacing counterterms is in contrast to the larger number of terms that one would need in order to correctly fit data from simulations with Wilson valence quarks on Wilson sea
quarks.  Because the Wilson action breaks chiral symmetry at $\c{O}(a)$, even for massless quarks, there will be terms proportional to all powers of the lattice spacing in the expression for the scattering length in Wilson $\chi$PT~\cite{Chen:noasqd,Rupak:2002sm,Aoki:2005mb}.  Moreover, such lattice spacing corrections begin at $\c{O}(a)$, rather than $\c{O}(a^2)$.  If one uses $\c{O}(a)$ improved Wilson quarks, then the leading discretization effects are of $\c{O}(a^2)$, as for staggered quarks; however, this does not remove the additional chiral symmetry-breaking operators.  
Another practical issue is whether or not one can perform simulations with
Wilson sea quarks that are light enough to be in the chiral regime.

%
%
%
%
%
%
%
%
%
%
%
%
\section{Discussion}\label{sec:summary}

Considerable progress has recently been made in fully
dynamical simulations of pion scattering in the $I=2$
channel~\cite{Yamazaki:2004qb,Beane:2005rj}. We have
considered $I=2$ scattering of pions composed of Ginsparg-Wilson quarks
on a staggered sea. We have calculated the scattering length in both this mixed action theory and
in continuum PQ$\chi$PT for theories with either 2 or $2+1$ dynamical
quarks. These expressions are necessary for the correct continuum and
chiral extrapolation of PQ and mixed action lattice data to the physical
pion mass.

Our formulae, Eqs.~\eqref{eq:2seaScattLength},~\eqref{eq:2P1seaScattLength}, not only provide the form for the mixed action scattering length, but also contain two predictions relevant to the recent work of Ref.~\cite{Beane:2005rj}.  Beane \emph{et.~al.}~calculated the $I=2$ $s-$wave $\pi\pi$ scattering length using domain wall valence quarks and staggered sea quarks, but used the continuum $\chi$PT expression to extrapolate to the physical quark masses.  In Figure~2 of Ref.~\cite{Beane:2005rj}, which plots  $m_\pi a_2^{(0)}$ versus $m_\pi / f_\pi$, the fit of the $\chi$PT expression to the lattice data overshoots the lightest pion mass point but fits the heavier two points quite well.  This is interesting because Eq.~\eqref{eq:2seaScattLength} predicts a \emph{known}, positive shift to $m_\pi a_2^{(0)}$ of size $\tilde\Delta_{PQ}^4/(768 f_\pi^4 \pi^3)$.  Accounting for this positive shift is equivalent to lowering the entire curve, and could therefore move the fit such that it goes between the data points.  This turns out, however, not to be the case.  In Ref.~\cite{Beane:2005rj}, the valence and sea quark masses are tuned to be equal, so $\tilde\D_{PQ}^2 = a^2 \D_I \simeq (446 \textrm{ MeV})^2$~\cite{Aubin:2004fs}.  Despite the large value of $\tilde\D_{PQ}$, the predicted shift is insignificant, being an order of magnitude less than the statistical error. In Table~\ref{t:ashifts}, we collect the predicted shifts to $m_\pi a_2^{(0)}$ at the three pion masses used in Ref.~\cite{Beane:2005rj}.    We also list the magnitude of the ratio of these predicted shifts to the leading contribution to the scattering length, which turn out to be small, lending confidence to the power counting we have used, Eq.~\eqref{eq:epsilons}.  
The other more important prediction is that there are no unknown corrections to the \CPT\ formula for the scattering length arising from lattice spacing corrections or partial quenching through the order $\c{O}(m_q^2)$, $\c{O}(a^2 m_q)$ and $\c{O}(a^4)$.  Therefore, to within statistical and systematic errors, the continuum \CPT\ expression used by Beane \emph{et.~al.}~to fit their numerical $\pi\pi$ scattering 
data~\cite{Beane:2005rj} receives no corrections through the 1-loop level.

\begin{table}
\caption{Predicted shifts to the scattering length computed in Ref.~\cite{Beane:2005rj} arising from finite lattice spacing effects in the mixed action theory.  The first two rows show the approximate values of $m_\pi$ and $f_\pi$  while the third shows $m_\pi a_2^{(0)}$ plus the statistical error calculated in \cite{Beane:2005rj}.  In the fourth row, we give the predicted shifts in the scattering length (times $m_\pi$) and, in the fifth row, we give the ratio of the predicted shift to the leading order contribution to the scattering length.}
\begin{tabular}{| c | c c c |}
\hline
 $m_\pi$ (MeV) & $294$ & $348$ & $484$   \\
 $f_\pi$ (MeV) & $145$ & $149$ & $158$ \\
$ m_\pi a_2^{(0)}$ 
& $-0.212 \pm 0.024$ 
& $-0.222 \pm 0.014$ 
& $-0.38 \pm 0.03$ \\
\hline
$\frac{\tilde\D_{PQ}^4}{768 \pi^3 f_\pi^4}$
 & $0.00374$ 
 & $0.00336$ 
 & $0.00266$  \\    
$\frac{\tilde\D_{PQ}^4}{6 (4\pi f_\pi m_\pi)^2}$
 & $0.0229$ 
 & $0.0155$ 
 & $0.00711$ \\
\hline
\end{tabular}
\label{t:ashifts}
\end{table}

The central result of this paper is that the appropriate way to extrapolate lattice $\pi\pi$ scattering data
is in terms of the lattice-physical pion mass and decay constant rather than in terms of the LO parameters which appear in the chiral Lagrangian.  When expressed in terms of the LO parameters, the scattering length depends upon 4 undetermined parameters, $l^\prime_{\pi\pi}$, $l^\prime_{PQ}$, $l^\prime_{a^2}$ and $C_\textrm{Mix}$.  In contrast, the scattering length expressed in terms of the lattice-physical parameters depends upon only one unknown parameter, $l_{\pi\pi}$, the same linear combination of Gasser-Leutwyler coefficients which contributes to the scattering length in continuum \CPT.

%
%
%
%
%
%
%
%
%
%
%
%
\begin{acknowledgments}
We would like to thank Maarten Golterman, Martin Savage, and Steve Sharpe for many useful discussions and helpful comments on the manuscript.  RV would like to thank Jack
Laiho for a helpful discussion and for pointing out a relevant reference.  And we would also would like to thank the referee for helpful comments and questions regarding the manuscript.  
JWC and AWL would like to thank the ECT* and the organizers
of {\it Effective theories in Nuclear physics and Lattice QCD}, Paulo Bedaque, Elisabetta Pallante, and Assumpta Parreno for
organizing a wonderful conference and providing a stimulating atmosphere
where we formulated the idea for this project. JWC thanks the INT at
the University of Washington for hospitality. DOC thanks the nuclear
theory group at the University of Washington for hospitality during
the completion of this work. JWC is supported by the National Science
Council of R.O.C.. DOC is supported in part by the U.S. DOE
under the grant DE-FG03-9ER40701. RV was supported under DOE grant
DE-FG02-96ER40956. AWL is supported under DOE grant DE-FG03-97ER41014.
\end{acknowledgments}

\bibliography{MasterBib}

\end{document}